\documentclass[%
twocolumn,
superscriptaddress,
nofootinbib,
 amsmath,amssymb,
 aps, prd
]{revtex4-2}

\newcommand\qone[1]{$q=1.{#1}$}
\newcommand\seo[0]{SEOBNRv4Tidal }
\newcommand\imr[0]{IMRPhenomD\_NRTidalv2 }

\usepackage{soul}
\usepackage{ulem}
\newcommand{\etal}[0]{\textit{et al}}

\usepackage{mathtools}
\usepackage{subcaption}
\usepackage{graphicx}
\usepackage{amsmath}

\newcommand{\WSU}{\affiliation{Department of Physics \& Astronomy,
	Washington State University, Pullman, Washington 99164, USA}}
\newcommand{\UNH}{\affiliation{Department of Physics \& Astronomy, University of New Hampshire, 9 Library Way, Durham NH 03824, USA}}
\newcommand{\TAPIR}{\affiliation{TAPIR, Walter Burke Institute for Theoretical Physics, MC 350-17, California Institute of Technology, Pasadena, California 91125, USA}}
\newcommand{\Cornell}{\affiliation{Cornell Center for Astrophysics and Planetary Science, Cornell University, Ithaca, New York, 14853, USA}}
\newcommand{\MPI}{\affiliation{Max Planck Institute for Gravitational Physics (Albert Einstein Institute), D-14467 Potsdam, Germany}}

\begin{document}

\title{Gravitational Waves from Binary Neutron Star Mergers with a Spectral Equation of State}
\author{Alexander Knight}\UNH
\author{Francois Foucart}\UNH
\author{Matthew D. Duez}\WSU
\author{Mike Boyle}\Cornell
\author{Lawrence E. Kidder}\Cornell
\author{Harald P. Pfeiffer}\MPI
\author{Mark A. Scheel}\TAPIR

\begin{abstract}
  In numerical simulations of binary neutron star systems, the equation of state of the dense neutron star matter is an important factor in determining both the physical realism and the numerical accuracy of the simulations.
  Some equations of state used in simulations are $C^2$ or smoother in the pressure/density relationship function, such as a polytropic equation of state, but may not have the flexibility to model stars or remnants of different masses while keeping their radii within known astrophysical constraints.
  Other equations of state, such as tabular or piece-wise polytropic, may be flexible enough to model additional physics and multiple stars' masses and radii within known constraints, but are not as smooth, resulting in additional numerical error.
  We will study  in this paper a recently developed family of equation of state, using a spectral expansion with sufficient free parameters to allow for a larger flexibility than current polytropic equations of state, and with sufficient smoothness to reduce numerical errors compared to tabulated or piece-wise polytropic equations of state.
  We perform simulations at three mass ratios with a common chirp mass, using two distinct spectral equations of state, and at multiple numerical resolutions.
  We evaluate the gravitational waves produced from these simulations, comparing the phase error between resolutions and equations of state, as well as with respect to analytical models.
  From our simulations we estimate that the phase difference at merger for binaries with a dimensionless weighted tidal deformability difference greater than $\Delta \tilde{\Lambda} = 55$ can be captured by the SpEC code for these equations of state.
\end{abstract}

\maketitle

\section{Introduction}

Great progress has been made in gravitational wave astrophysics starting with the binary black hole (BBH) merger event GW150914~\cite{Abbott_2016} and the detection of the binary neutron star merger event GW170817 by the LIGO/Virgo/KAGRA collaboration~\cite{Abbott_201712}. 
Additional mergers of black hole-neutron star (BHNS) and binary neutron star (BNS) have been observed, such as GW190425~\cite{Abbott_202003}, GW200115~\cite{Abbott_202106}, and GW191219~\cite{Abbott_202106b}.
We have also observed merger events between objects with properties outside of what was previously expected, such as GW190814~\cite{Abbott_202006}, which has either the heaviest NS or lightest BH detected so far, with a mass of $2.50-2.67 M_\odot$, partnered with a $22.2-24.3 M_\odot$ BH. 
Gravitational waves from these mergers carry information about the system's chirp mass, as well as the angular momentum and mass of each compact object. 
In the presence of a neutron star, they also provide information on how matter behaves at densities higher than nuclear saturation density.

To interpret this information, we need to quickly generate millions of theoretical gravitational wave signals.
Even with modern methods and technologies, however, creating simulations that sample the available parameter space densely enough to allow us to perform parameter estimation on observed binaries remains prohibitive in time and computational cost.
As a result, simulations are instead used to test and/or train analytical models.
While analytical models have high accuracy during the inspiral of a binary even without including any information from merger simulations, they can become inaccurate in the last few orbits before merger -- which also happens to be when the impact of the finite size of a neutron star is most noticeable on the waveform.
Numerical relativity simulations are thus used to test or calibrate these models for better accuracy during these critical stages of a merger event.
As the accuracy of these analytical models (or, at the very least, our ability to test said accuracy) is then dependent on the accuracy of merger simulations, a reduction of potential numerical errors below the expected accuracy of current and future gravitational wave detectors is highly desirable.
For GW170817, the only event so far to put meaningful constraints on the equation of state (EoS) of dense matter, modeling errors were most likely unimportant~\cite{Abbott_201712}. 
However, with planned improvements to the sensitivity of current detectors and the construction of new observatories, the frequency of gravitational wave detections will increase, and modeling errors will potentially become an important source of uncertainty.

In the case of BNS mergers, high accuracy simulations are still relatively new, and even the best simulations have a lower accuracy and a higher computational cost than for BBH mergers.
BNS mergers require the evolution of the equations of general relativistic hydrodynamics in addition to Einstein's equations. 
Beyond the cost of evolving this additional system of equations, the presence of surface discontinuities and/or shocks in the fluid makes it impossible to use for neutron star mergers the high-order methods that have allowed BBH simulations to produce high-accuracy waveforms at a reasonable computational cost, or at least impossible to consistently use such methods across the entire computational domain.

There are at least two ways to limit the unphysical impacts of discontinuities in the evolution of the fluid. 
One is the use of numerical methods capable of maintaining high-order accuracy while capturing shocks and surfaces. 
Methods demonstrating third order convergence (for beta-equilibrium EoS)~\cite{Radice_2014} and fourth order convergence (for piecewise polytropes)~\cite{Doulis:2022vkx} in the phase of the gravitational wave signal have been published so far.
The other is to improve the smoothness of the EoS used to describe high-density matter~\cite{Foucart_201911,Raithel_202207}.
Smoother EoS typically lead to lower truncation errors. 
A simple $\Gamma=2$ polytropic EoS, for example, is much preferable in terms of numerical accuracy to the more complex piecewise-polytropic EoS, or to the tabulated, composition and temperature dependent EoS needed in simulations that evolve neutrinos or that aim to capture changes in the fluid composition.
There is however a significant trade-off, in that a simple polytropic EoS matching the desired mass and radius of a specific neutron star will often not be consistent with known physical constraints, for example the maximum mass of neutron stars, or the radius of neutron stars of different masses. 
Additionally, these simple EoS are much farther from satisfying constraints on the nuclear EoS derived from theoretical and experimental nuclear physics results.
With constraints on the neutron star from studies such as~\cite{Abbott_2018,Raaijmakers_2020,Raaijmakers_2021,Carson_201902,Carson_201907}, a smooth EoS that is more consistent with at least the known physical constraints on the macroscopic properties of cold neutron stars (mass-radius relation, maximum mass) is desirable.

For these reasons, this paper will be evaluating our ability to perform high-accuracy simulations with a relatively new EoS utilizing a spectral representation of the nuclear EoS, which results in a smoother EoS with more flexibility in setting the pressure $P(\rho)$ and specific internal energy $\epsilon(\rho)$ for dense matter than in simple polytropes. 
We note that while this does allow us to construct an EoS that more closely matches any chosen set of constraints on the properties of cold matter in beta-equilibrium, the temperature dependence of our EoS remains extremely simplified, and it does not include any composition dependence~\cite{Foucart_201911}. 
A method to expand a cold, beta-equilibrium EoS with a more physically motivated temperature and composition dependence has been proposed in~\cite{Raithel_202209,Raithel_2019}, but is not currently implemented in the SpEC code used in this work. 
Our ability to accurately evolved these spectral EoS was first tested in~\cite{Foucart_201911} over short simulations; this manuscript presents our first full inspiral-merger simulations using these EoS. 
As we are in particular interested in estimating our ability to capture tidal effects in waveforms, this manuscript includes simulations of systems with identical neutron star masses but performed with two EoS with $\sim 20\%$ differences in their dimensionless tidal deformability ($\sim 570$ vs $\sim 710$ for GW170817-like systems). 
We will show that the phase difference between the resulting waveforms at merger is resolved in our simulations.

In addition to our study of this spectral EoS, we will in this manuscript evaluate the performance of a new time stepping method for SpEC evolutions, where the evolution of Einstein's equations is permitted to take smaller steps than the evolution of the fluid equations.
For the SpEC code, this offers reduced computational cost, as these two systems of equations are evolved on different numerical grids, with different time step constraints.
The fluid equations are typically more costly to solve in each individual time step while Einstein's equations often require a shorter step to reach a desired accuracy.
Here, we demonstrate that uncoupling these time stepping systems results in simulations equivalent (within expected numerical errors) to those obtained when Einstein's equations and the equations of hydrodynamics use the same time stepping algorithm.

We simulate six distinct physical configurations, using two different EoS each evolved at three mass ratios. 
For two of these cases, we also use both time stepping methods.
We then extrapolate the gravitational waves to future null infinity, and compare these gravitational waves to analytical models. All waveforms presented here will will become public as part of the next data release by the SxS collaboration.

\section{Methods}

\subsection{Evolution}
For the simulations presented in this manuscript, we use the Spectral Einstein Code (SpEC)~\cite{spec}, which evolves Einstein's equations using the Generalized Harmonic formalism~\cite{Lindblom_2006} on a pseudospectral grid, with p-type adaptive mesh refinement~\cite{Szilagyi_2014}.
The general relativistic hydrodynamical equations are evolved on a separate grid using the high-order shock capturing scheme described in Radice and Rezzolla (2012)~\cite{Radice_2012}. 
This scheme uses the fifth-order, shock capturing MP5 reconstruction to interpolate from cell centers to cell-faces, and a Roe solver to calculate numerical fluxes at cell faces. 
It has been shown to result in third-order convergence of the solution when used in neutron star merger simulations~\cite{Radice_2012}.
    
Einstein's equations and the fluid equations are both evolved in time using a third-order Runge-Kutta algorithm.
In the algorithm previously used in SpEC, both systems of equations use the same time step, chosen to meet a target time discretization error (see Appendix A, section 3 of~\cite{Foucart_2013}).
Coupling source terms are communicated between the grids at the end of each time step. 
Linear extrapolation from the current step and previous steps is used to determine the values of the source terms during the intermediate steps of the Runge-Kutta algorithm. 

In this manuscript, we also use for the first time an algorithm where the evolution of Einstein's equations is allowed to use a smaller time step than the evolution of the fluid. 
In that algorithm, the fluid equations take a time step $\Delta t = \alpha_{CFL} \min{\left(\frac{\Delta x}{c_{\rm max}}\right)}$, with $\Delta x$ the grid spacing, $c_{\rm max}$ the maximum characteristic speed of the fluid equations in grid coordinates in that cell (in absolute value), and $\alpha_{\rm CFL}=0.25$ a constant chosen to maintain stability of the evolution. 
The time step used for Einstein's equations is chosen in the same way as in our standard algorithm, with the minor modification that we require each time step of the fluid evolution to be an integer number of time steps of the metric evolution. 
In this new method, source terms are communicated between the two grids at the end of each time step of the fluid evolution. 
This results in less frequent communication than in our previous algorithm, and in a lower number of time steps for the fluid equations. 
At intermediate times, we again use extrapolation from previous time steps to calculate the source terms. The order of extrapolation used in this algorithm is freely specifiable. So far, we found no significant impact on the accuracy of our simulations as long as we use at least first order extrapolation.
In the rest of this manuscript, we will refer to the simulations using the same time step on both grids as {\it Shared Time Step} (ShTS), and the simulations using a different time step on each grid as {\it Split Time Step} (SpTS).

    Interpolating from the spectral grid to the finite-difference grid to communicate source terms is done by refining the spectral grid by approximately a factor of 3 in each dimension, and then using third-order polynomial interpolation from the colocation points in the refined spectral grid to the finite-difference grid.
    Interpolation from the finite-difference grid to the spectral grid uses fifth-order polynomial interpolation, limited so that interpolation does not create any new extremum in the fluid variables.

    The pseudospectral and finite difference grids both rotate and contract to follow the binary system.  
    The finite-difference grid is rescaled  when the grid spacing decreases by a factor of 0.8 in the inertial frame, in order to keep a consistent resolution during all phases of the evolution.
    The finite-difference grid removes subdomains where no significant matter ($\text{max}(\rho) < 6.2\times10^{9}\frac{g}{cm^3}$) 
is located, and adds back subdomains as higher density matter flows close to the boundary of the removed subdomains over the course of the simulation.
    These two methods result in a reduced computational cost for our simulations.

    For a full explanation on SpEC's methods for the evolution of the hydrodynamical and general relativistic grids, we refer the reader to Duez \etal (2008)~\cite{Duez_2008}, as well as appendix A of Foucart \etal (2013)~\cite{Foucart_2013}.
    We will limit ourselves here to a brief discussion of the methods most relevant to the use of spectral EoS in simulations aiming to produce high-accuracy numerical waveforms.

    We define the neutron star matter as a perfect fluid with stress-energy tensor
    \begin{equation}
      T^{\mu \nu} = (\rho + u + P)u^\mu u ^\nu + P g^{\nu \mu}.
    \end{equation}
    In this equation, we have the pressure $P$, baryon density $\rho$, internal energy density $u$, 4-velocity $u^\mu$, and the inverse metric $g^{\mu \nu}$.
    The evolution equations are derived from the conservation of baryon number 
\footnote{The baryon density is defined as $\rho=m_b n$, with $n$ the baryon number density and $m_b$ an arbitrarily chosen reference mass for baryons; accordingly, our evolution equation represents conservation of baryon number, not conservation of mass.}
    \begin{equation}
      \nabla_\mu (\rho u^\mu) = 0
    \end{equation}
    and the energy-momentum conservation
    \begin{equation}
      \nabla_\mu T^{\mu \nu}=0,
    \end{equation}
    which give 5 equations for 6 independent variable (e.g. $\rho,u,P$ and 3 independent components of the velocity).
    We close the system of equations with an EoS, 
which introduces two functions $P(\rho,T)$ and $u(\rho,T)$ (defined below), with temperature $T$.
    While such an EoS introduces a new variable, the two additional equations are sufficient to close the system of equations.
    
    Practically, we evolve the 'conserved' variables
    \begin{eqnarray}
      \rho_* &=& -\sqrt{\gamma}n_\mu u^\mu \rho,\\
      \tau &=& \sqrt{\gamma} n_\mu n_\nu T^{\mu \nu}-\rho_*,\\
      S_k &=& -\sqrt{\gamma} n_\mu T^{\mu}_k,
    \end{eqnarray}
    where $\gamma$ is the determinant of the spacial metric $\gamma_{ij}=g_{ij}+n_i n_j$, and $n^\mu$ is the future directed unit normal to a constant time slice.
    The integral of $\rho_*$ (``total baryonic mass'') is conserved over the entire domain, up to losses at the domain boundary.
    Recovering the ``primitive'' variables $(\rho_0,T,u_i)$ from the conserved variables requires multi-dimensional root-finding. 
    We follow the 2D root-finding method of Noble {\it et al}~\cite{Noble2006}, with corrections in low-density and high-velocity regions where due to numerical errors in the conservative variables the inversion may not be possible~\cite{Foucart_2013}.
    The evolution equations are written in ``conservative'' form, i.e. as a set of five coupled equations of the form
    \begin{equation}
      \frac{\partial \textbf{F}^0(\textbf{u})}{\partial t} +\sum_{i=1}^{3} \frac{\partial \textbf{F}^i(\textbf{u})}{\partial x^i} = \textbf{S(u)}
    \end{equation}
    with the primitive variables $\textbf{u}$, vector of conserved variables $\textbf{F}^0(\textbf{u})$, fluxes $\textbf{F}^i$, and source terms $\textbf{S(u)}$.
    These fluxes and source terms are calculated at cell centers, and the fluxes (as well as the physical variables $\rho_0$, $T$, and $u_i$) are interpolated to cell faces. 
    The calculation of the divergence of the fluxes from the values of $\textbf{u}$ at cell centers follows the previously mentioned method of Radice \& Rezzolla~\cite{Radice_2012}.
    
    \subsection{Numerical Implementation of Spectral Equation of State}

    The EoS used in these simulations to describe matter inside the neutron star was developed in Lindblom (2010)~\cite{Lindblom_2010}, and modified in Foucart \etal (2019)~\cite{Foucart_201911} for computational use. 
As already mentioned, the choice of EoS in numerical simulations is often a trade-off between the ability to capture more physics and a wider range of possible models on one side, and the numerical accuracy of the simulations on the other side. 
Spectral EoS are smoother than both tabulated and piecewise polytropic EoS, which results in higher accuracy simulations (as will be seen in the results section). 
On the other hand, our existing spectral EoS are limited to matter in beta-equilibrium, use an extremely simplified model for the thermal pressure, and are not suitable for coupling to neutrino evolution. 
We consider this a reasonable trade-off when attempting to generate high-accuracy gravitational wave signals from the inspiral, merger, and early post-merger evolution of BNS and BHNS binaries, but acknowledge that spectral EoS would be a poor choice for simulations attempting e.g. to model the outcome of r-process nucleosynthesis in mergers. 
Single polytropic EoS, for their part, lead to higher accuracy simulations than more complex spectral EoS at a given resolution\footnote{Or at least than the spectral EoS used in this manuscript; single polytropic EoS are themselves a subset of the spectral models, but one that does not allow much flexibility on the functional form of the EOS.}. 
However, their use to simulate asymmetric binaries and/or the merger and post-merger phase of the evolution of a binary may be problematic. 
Indeed, while it is possible to construct a single polytropic EoS for which a neutron star of a given mass has the desired radius, or the desired tidal deformability, it is typically difficult to do this for two neutron stars of distinct mass, or to make sure that the EoS at the same time support massive neutron stars $M_{\rm NS}\gtrsim 2M_\odot$.

    In this section, we give a reduced explanation of the theory of spectral EOS. 
    A full explanation can be found in the previous work by Foucart \etal (2019)~\cite{Foucart_201911} and Lindblom (2010)~\cite{Lindblom_2010}. 

    We choose the spectral expansion as in Foucart \etal (2019)~\cite{Foucart_201911} by writing the pressure $P$ and specific internal energy $\epsilon$ as

    \begin{equation}
  P(x,T)=
  \begin{dcases}
    P_0 \exp\bigg(\Gamma_0 x + \eta_2 \frac{x^3}{3} + \eta_3 \frac{x^4}{4}\bigg) + \rho T & x>0\\
    P_0 \exp(\Gamma_0 x) + \rho T & x\leq 0
  \end{dcases}
    \end{equation}

    and

    \begin{equation}
  \epsilon(x,T)=
  \begin{dcases}
    \epsilon_0 + \int_0^x d\xi \frac{P(\xi,0)}{\rho_0}e^{-\xi}+\frac{T}{\Gamma_{th}-1} & x>0\\
    \frac{P(x,0)}{\rho(\Gamma_0-1)}+\frac{T}{\Gamma_{th}-1} & x \leq 0
  \end{dcases}
    \end{equation}

    with some reference density $\rho_0$, reference adiabatic index $\Gamma_0$, reference pressure $P_0$, temperature $T$, and where we define $x = \log\bigg(\frac{\rho}{\rho_0}\bigg)$. 
    We note that despite its name, used here to match standard conventions, $T$ does not scale linearly with the physical temperature of the fluid; it is simply defined so that the thermal pressure is $P_{\rm thermal} = \rho T$.

\begin{figure}
  \includegraphics[width=\columnwidth]{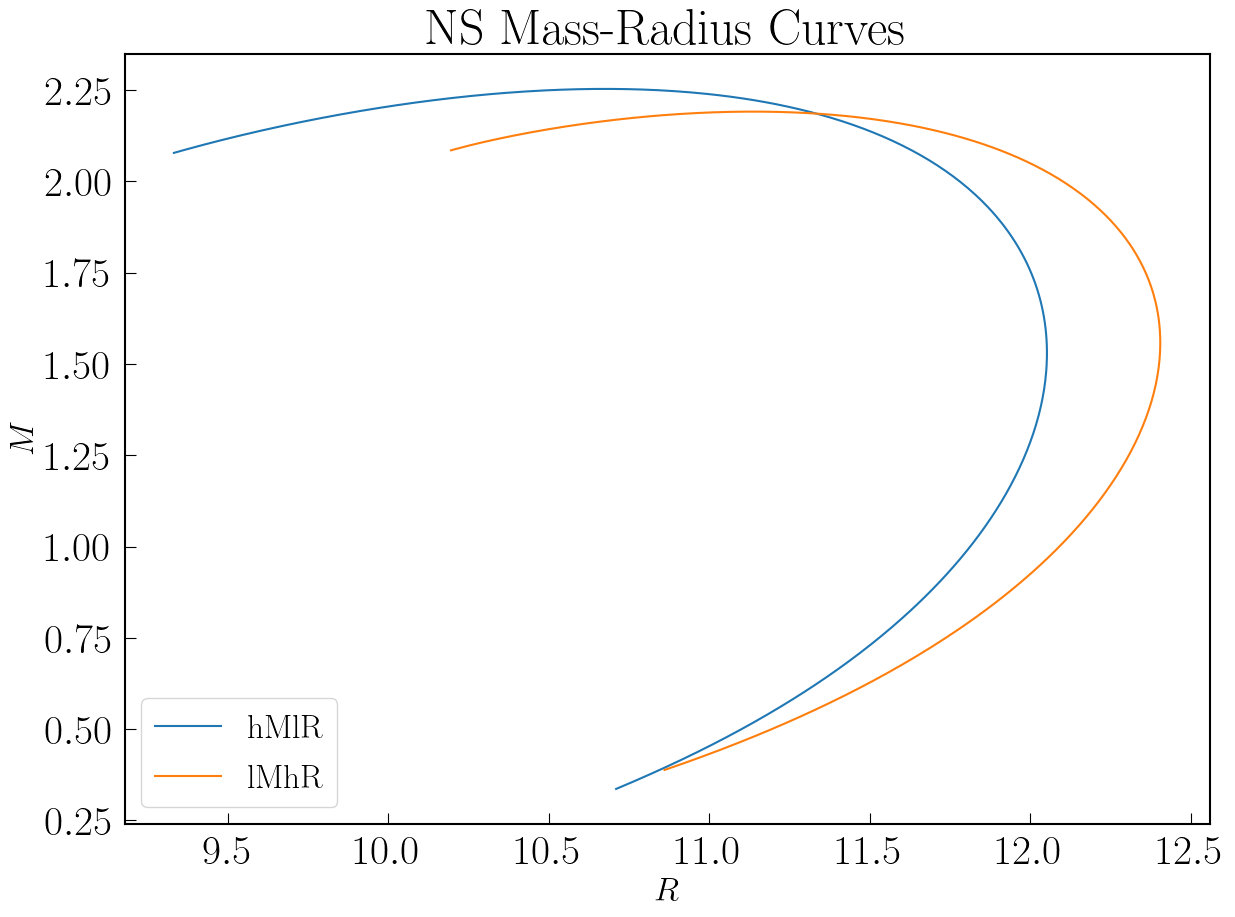}
  \caption{Mass-radius curves for our EoS parameter sets.
           We shorthand these EoS by their behavior on this plot, with the blue curve having a higher maximum mass, but lower maximum radius, and as such abbreviated as hMlR.
           We shorthand the orange curve to lMhR in a similar fashion.}
  \label{fig:mass_radius_curves}
\end{figure}

    These equations give free parameters of $\eta_2$, $\eta_3$, $\rho_0$, $P_0$, and $\Gamma_{0}$.
    Many possible sets of values for these would result in sound waves in dense matter moving at superluminal velocities and/or behavior that does not conform to known nuclear physics. 
    In Foucart \etal (2019)~\cite{Foucart_201911}, a Marko-Chain Monte Carlo method was used to determine values of these parameters resulting in causal EoS, and values of the pressure at high-density within the range of values currently allowed by nuclear physics.
We also found that the choice $\Gamma_0=2$ led to higher accuracy than lower values of $\Gamma_0$, possibly due to the simple behavior of the density close to the surface for that choice of $\Gamma_0$. 
We make this choice here as well, even though this results in values of the pressure and internal energy at $\rho \ll \rho_0$ that are inconsistent with the known behavior of dense neutron rich matter at low density. 
This is reasonable for our purpose here because the gravitational wave signal is mostly sensitive to the EoS at high density.
    From~\cite{Foucart_201911}, we choose two EoS with free parameters shown in table \ref{tab:eos_parameters}, with mass-radius curves shown in figure \ref{fig:mass_radius_curves}.
    The first EoS parameter set displays a higher maximum mass and lower maximum radius (hMlR) while the second has a lower maximum mass and a higher maximum radius (lMhR).
    The hMlR EoS gives neutron stars with a maximum Schwartzschild radius of 12.05 km and maximum baryonic mass of 2.719 $M_\odot$, while the lMhR EoS gives neutron stars with a maximum Schwartzschild radius of 12.41 km and maximum baryonic mass of 2.191 $M_\odot$.

  \subsection{Initial conditions}

  \begin{table*}[t]
    \centering
    \begin{tabular}{||c||c|c|c|c|c|c|}
      \hline
      EoS & $\Gamma_0$ & $\eta_2$ & $\eta_3$ & $\Gamma_\text{th}$ & $\rho_0\frac{\text{g}}{\text{cm}^3}$ & $P_0\frac{\text{dyn}}{\text{cm}^2}$ \\
      \hline \hline
      hMlR & 2.0 & 0.45872 & -0.114849 & 1.75 & $5.07405\times 10^{13}$ & $1.637887436\times10^{35}$ \\
      \hline
      lMhR & 2.0 & 0.435096 & -0.111447 & 1.75 & $3.33533\times 10^{13}$ & $5.8481\times 10^{34}$ \\
      \hline
    \end{tabular}
    \caption{Spectral EoS parameters used in this manuscript. Mass-radius curves for these EoS parameter sets can be seen in figure \protect{\ref{fig:mass_radius_curves}}.
    }
    \label{tab:eos_parameters}
  \end{table*}
    For both the ShTS and SpTS time stepping algorithm, we evolve binary neutron star systems with the same chirp mass $M_{\rm chirp}=1.18M_\odot$, chosen to match the chirp mass of GW170817. 
    Our chosen configurations for the ShTS method consist of two systems, with mass ratios of \qone1 and \qone2 and the hMlR EoS, separated by a distance of $53.1$km, and no initial neutron star spin.
    The SpTS simulations have mass ratios of \qone0, \qone1, and \qone2, separated by $54.6$km, with no initial spin, and are performed for both EoS.
    We construct out initial data utilizing our SPELLS code~\cite{Pfeiffer_2003} adapted for binary neutron star systems~\cite{Foucart_2008,Haas_2016}, which generates a binary system in quasi-circular orbit.
    From this, we iteratively adjust the initial angular and radial velocity of the neutron stars to reduce the initial eccentricity of the orbits to $\lesssim0.002$ utilizing the methods of Pfeiffer (2007)~\cite{Pfeiffer_2007}.
    With these methods, and initial separation distance, we reach merger at about 10.5 orbits for the ShTS and about 11.5 orbits for the SpTS.
    Parameters for all simulations can be found in table \ref{tab:bns_parameters}.

 With these EoS, we have two systems with a mass averaged dimensionless tidal deformability $\tilde{\Lambda}\approx 570$ for the hMlR EoS and $\tilde{\Lambda}\approx 710$ for the lMhR EoS.
 For the hMlR EoS, the dimensionless tidal deformabilities ranges are $\Lambda_1=318.655-588.389$ and $\Lambda_2=588.389-990.603$, and the lMhR EoS ranges from $\Lambda_1=529.503-713.648$ and $\Lambda_2=713.648-1224.13$.
 These values rest comfortably within the 90\% probability region of low spin systems (and very close to the 50\% region in some cases) for the $\Lambda_1$ and $\Lambda_2$ relationship from the LIGO and Virgo constraints~\cite{Abbott_201710}.

\begin{table*}[t]
  \centering
  \begin{tabular}{||c||c|c||c|c|c|c|c|c|c|c|c||}
    \hline
    Time Step & $q$ & EoS & $M_1 (M_\odot)$ & $M_2 (M_\odot)$ & $R_1$ (km) & $R_2$ (km) & $d$ (km) & $\Omega_0$ (kHz) & $\Lambda_1$ & $\Lambda_2$ & $\tilde{\Lambda}$\\
    \hline \hline
    ShTS & 1.1 & hMlR & 1.4267 & 1.2970 & 12.042 & 12.004 & 53.100 & 1.41243 & 424.209 & 766.046 & 573.011\\
    \hline
    ShTS & 1.2 & hMlR & 1.4910 & 1.24254 & 12.050 & 11.980 & 53.100 & 1.41454 & 318.655 & 989.482 & 572.203\\
    \hline
    \hline
    SpTS & 1.0 & hMlR & 1.3600 & 1.3600 & 12.024 & 12.024& 54.635 & 1.35516 & 588.389 & 588.389 & 588.389\\
    \hline
    SpTS & 1.1 & hMlR & 1.4268 & 1.2971 & 12.042 & 12.004 & 54.641 & 1.35649 & 424.339 & 766.544 & 573.289 \\
    \hline
    SpTS & 1.2 & hMlR & 1.4911 & 1.2426 & 12.050 & 11.980 & 54.645 & 1.35768 & 318.739 & 990.603 & 572.672\\
    \hline
    SpTS & 1.0 & lMhR & 1.3600 & 1.3600 & 12.363 & 12.363 & 54.636 & 1.35516 & 713.648 & 713.648 & 713.648\\
    \hline
    SpTS & 1.1 & lMhR & 1.4268 & 1.2971 & 12.386 & 12.334 & 54.639 & 1.35613 & 529.503 & 950.249 & 712.685\\
    \hline
    SpTS & 1.2 & lMhR & 1.4911 & 1.2426 & 12.400 & 12.302 & 54.644 & 1.35803 & 398.868 & 1224.13 & 710.935\\
    \hline
  \end{tabular}

  \caption{
    From left to right, we have the time stepping method, mass ratio $q$, the EoS (see table \protect{\ref{tab:eos_parameters}}), ADM mass in units of solar masses $M_1$ and $M_2$, areal radii in km $R_1$ and $R_2$, initial separation in km $d$, initial angular velocity $\Omega_0$ in Hz, dimensionless tidal deformability $\Lambda_1$ and $\Lambda_2$, and the weighted average tidal deformability $\tilde{\Lambda}$. 
}
  \label{tab:bns_parameters}
\end{table*}

   \subsection{Domain/Grid Setup}

    The initial finite-difference hydrodynamical domain construction consists of a rectangular, bar-shaped Cartesian grid space with the neutron stars located at each end.
    We have three resolutions for the \qone1 and \qone2 hMlR ShTS and the \qone0 lMhR SpTS simulations.
    For the \qone1 and \qone2 simulations, we have grid spacings of $\Delta x_{\text{FD}} = 298$m, 239m, 191m with number of grid points along each dimension of $(369\times185\times185)$, $(457\times229\times229)$, and $(577\times289\times289)$, respectively.
    The \qone0 simulation has grid spacings of $\Delta x_{\text{FD}}=273$m, 218m, and 174m, with grid points of $(401\times201\times201)$, $(505\times253\times253)$, and $(649\times325\times325)$, respectively.
    For future ease, we will refer to the lowest resolution of each simulation set as Lev0, the middle Lev1, and the highest resolution Lev2.
    The SpTS hMlR EoS simulations were run at the Lev1 resolution, while the SpTS lMhR \qone1 and \qone2 were run at the Lev1 and Lev2 resolutions. In all SpTS simulations, the grid spacing is chosen so that we have $N=(72,90,112)$ grid points across the diameter of the neutron stars at (Lev0,Lev1,Lev2), averaging over both stars at $t=0$\footnote{Note that our initial data is in a coordinate system close to isotropic coordinates, in which the neutron star radius is smaller than in Schwarzschild coordinates. Hence $N\times \Delta x \neq 2R_{\rm NS}$ if $R_{\rm NS}$ is the areal (Schwarzschild) radius used, by convention, in our description of the neutron stars.}.

    This domain is divided into 8 equal segments along the shorter axes, and 16 segments along the long axis, resulting in 512 subdomains of equal size.
    The spectral grid construction consists of a ball and 5 shells covering each of the neutron stars.
    A set of distorted cylinders, with the rotational axis along the line between the neutron stars, connects the sets of ball and shells around the neutron stars.
    These cylinders also connect to 12 shells covering the outer regions, which are centered on the center of mass of the system.
    After merger, the area interior to the outer shells is replaced by distorted cubic subdomains.
    We refer the reader to Foucart {\it et al} (2013)~\cite{Foucart_2013} and Szil\'agyi (2014)~\cite{Szilagyi_2014} for a more detailed explanation and graphics of the pseudospectral grid construction.
    The number of basis functions used in each subdomain is adaptively chosen to reach a user defined maximum error, which is estimated from the spectral coefficients of the evolved variables.
    The user-defined accuracy on the pseudospectral grid is chosen such that it scales as $(\Delta x_{\text{FD}}^0)^5$, and as such errors on this grid converge faster than those on the finite-difference hydrodynamical grid.

  \subsection{Waveform Extrapolation}

    The simulations evolve through inspiral, plunge, and merger, then continue until the peak of the gravitational waves resulting from the merger event progress past the outer edge of the pseudospectral grid at a radius of 2047.5 km for the ShTS simulations and 2074.7 km for the SpTS simulations.

    The method to extrapolate the gravitational wave signal to null infinity from the metric at finite radii follows the procedure outlined by Boyle \& Mroue (2009)~\cite{Boyle_2009}.
    The Newman-Penrose scalar $\Psi_4$ and metric perturbation $h$ are estimated on spheres of constant inertial radii and decomposed into spin=-2 spherical harmonics components.
    At 24 radii from $R_i=211.2$ km to $R=2,015.6$ km equidistant in $1/R$, we compute a retarded time $t_\text{ret}(t,R_i)$ to approximate the travel time for the wave from the merging neutron stars to $R_i$.
    From here, we fit the ansatz
    \begin{align}
      A_{lm}(t_{\text{ret}},r) &= \sum_{j=0}^{N} A_{lm,j}(t_{\text{ret}})r^{-j} \\
      \phi_{lm}(t_{\text{ret}},r) &= \sum_{j=0}^N \phi_{lm,j}(t_{\text{ret}})r^{-j}
    \end{align}
    to the amplitude $A_{lm}$ and phase $\phi_{lm}$ of the $(l,m)$ component of the spherical harmonic decomposition of the gravitational wave at fixed retarded times.
    We then estimate the $(l,m)$ mode at infinite radius to be $A_{lm,1}e^{i\phi_{lm,0}}$.
    
\section{Results}

\subsection{Error Analysis}

As in Foucart \etal (2019)~\cite{Foucart_201911} and Foucart \etal (2021)~\cite{Foucart_2021}, we present and utilize a standard method of error estimation that likely overestimates the potential errors.
We take into account three potential sources of error in our simulations: finite resolution of our computational domain, extrapolation of the gravitational wave to infinity, and mass lost during the simulation at the boundaries.
For a more detailed overview of how these error sources are evaluated, we recommend Foucart (2019)~\cite{Foucart_201911}, but we will review the fundamentals here, and show the error estimates for the (2,2) mode of the extrapolated waveforms.

We estimate the errors due to finite resolution by comparing the three resolutions, Lev0 (low), Lev1 (medium), and Lev2 (high).
First, we use the phases of the Lev0 and Lev2 simulations in a Richardson extrapolation to infinite resolution, assuming a 2nd-order convergence. 
We then take the difference between the Lev2 waveform and the extrapolated waveform as a first estimate of the numerical error.
We repeat this process on the Lev2 and Lev1 resolutions, and keep the worse of these two error estimates.
We note that this is typically conservative because the methods used within the SpEC code converge at 3rd order or better. However, the hybrid spectral/finite volume methods utilized by SpEC causes different errors to dominate at different phases of the simulation.
As a result, when considering only two resolutions, it is not uncommon for multiple sources of error to cancel each other during a simulation. In our experience so far comparing numerical simulations with different numerical setups, comparing with results from other collaborations, and in case were additional higher accuracy simulations were performed, using three resolutions and considering the worse error estimate when comparing the simulations pairwise offers a comfortable buffer against this issue.

The errors from extrapolation to null infinity are estimated by comparing the phase difference between the 2nd and 3rd order extrapolation in $r^{-1}$, between $t=0$ and $t_{\text{peak}}$, where the (2,2) mode of the waveform reaches its maximum amplitude.
The maximum phase difference is conservatively chosen to be the associated error. 
This error is typically smaller than the finite resolution error, except at the very beginning of a simulation.

Finally, mass lost during the evolution results in gravitational waves emitted from a system different from the initially intended system.
Here, we use an estimate of the resulting error in the phase of the waveform derived in Boyle (2007)~\cite{Boyle_2007}.
In our simulations, mass loss was minimal, and resulted in a negligible phase error comparative to the error from finite resolution.
The $q=1.1$ ShTS simulation lost approximately $5.42\times10^{-6} M_\odot$ while the $q=1.2$ simulation lost approximately $8.71\times10^{-6} M_\odot$.
The lMhR \qone0 SpTS simulation lost $1.52\times10^{-5} M_\odot$.
The other SpTS simulations were performed at either 1 or 2 resolutions, as opposed to the 3 required for the previous error analysis.
The simulations with 2 resolutions demonstrated similar phase difference between resolutions as the \qone0 lMhR SpTS.

\begin{figure*}
  \includegraphics[width=0.32\linewidth]{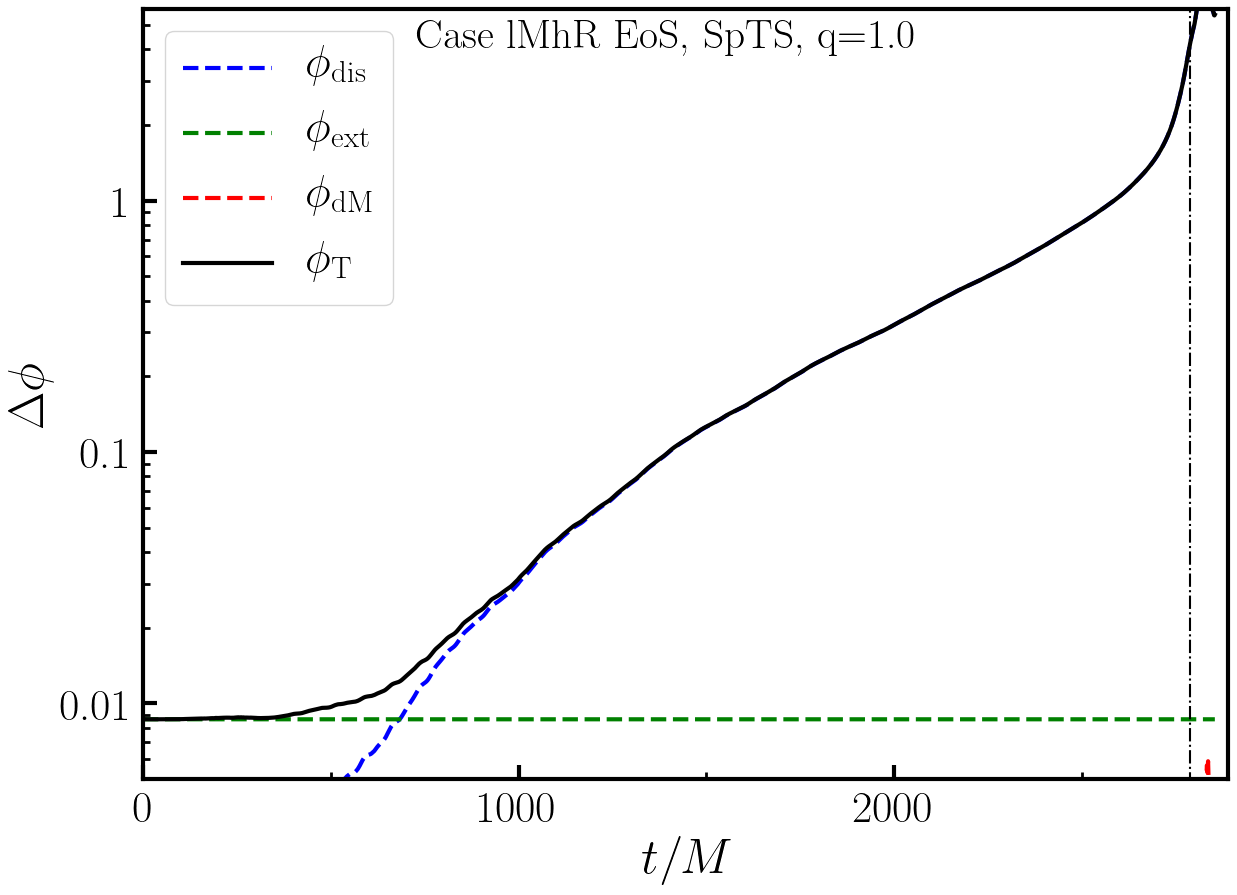}
  \includegraphics[width=0.32\linewidth]{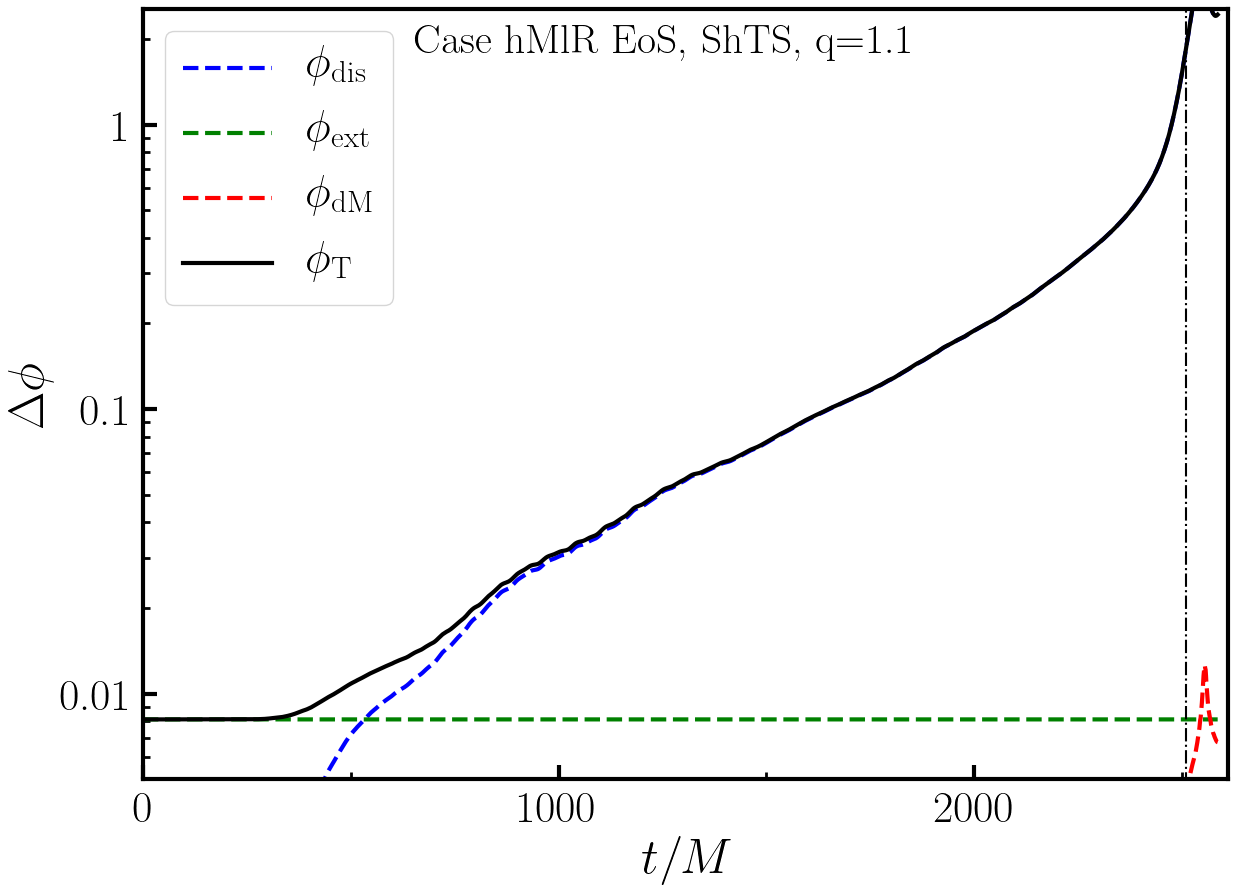}
  \includegraphics[width=0.32\linewidth]{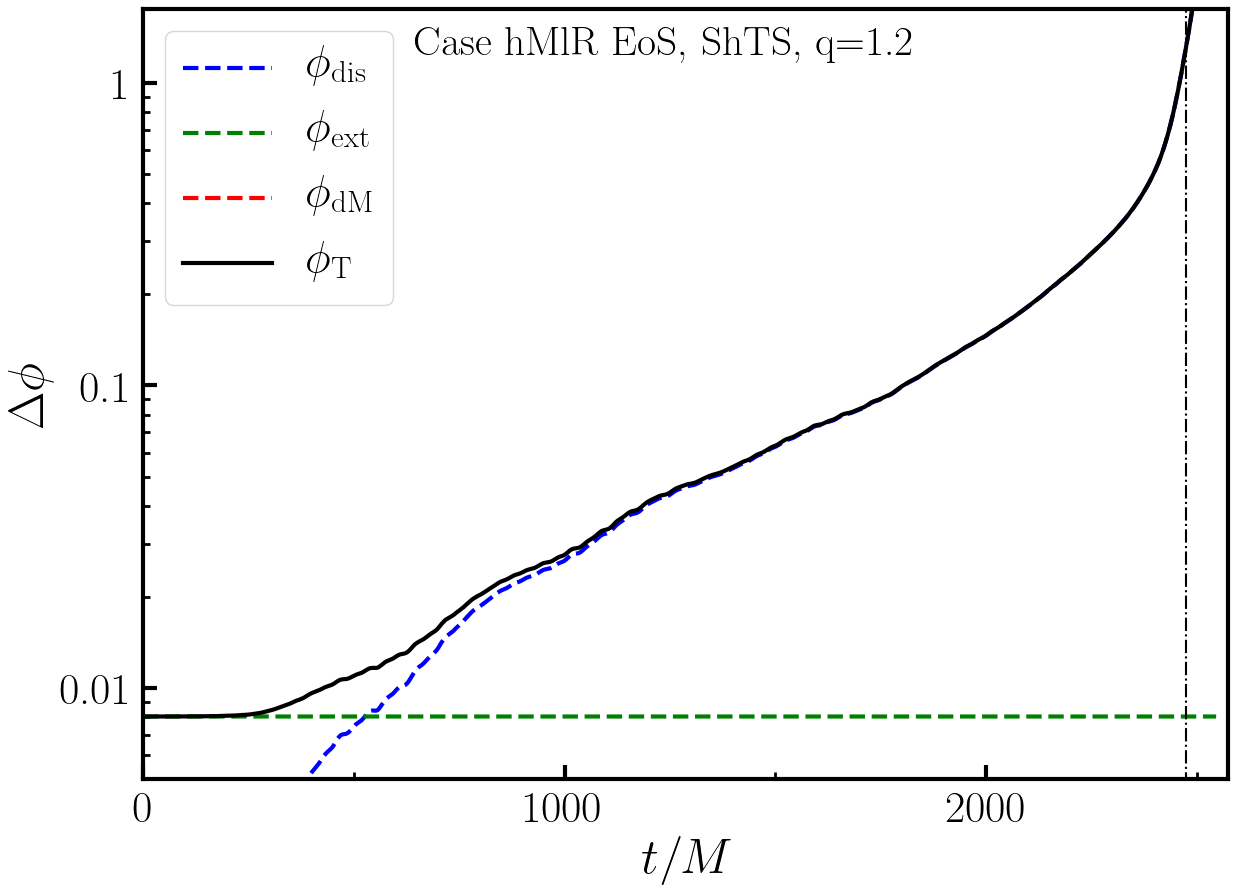}
  \caption{Phase error estimates for the (2,2) mode of the $q=1.0$ SpTS, \qone1 ShTS, and \qone2 ShTS binaries, broken down by sources: finite resolution ($\phi_{\text{dis}}$), extrapolation error ($\phi_{\text{ex}}$), and error from mass loss ($\phi_{\text{dM}}$), with the vertical dashed line indicating the peak of the gravitational wave amplitude (merger).
  }
  \label{fig:extraperror}
\end{figure*}

In figure \ref{fig:extraperror}, we can see the three sources of error, as well as the total error $\phi_T$.
In all of these figures, clearly the finite resolution error ($\phi_{\text{dis}}$) of the simulation dominates $\phi_T$ from a few hundred $t/M$ after the start of the simulation to past merger.
At the time of merger, the hMlR ShTS \qone1 and \qone 2 simulations have approximately 1-2 radians of phase error, but the longer lMhR SpTS \qone0 simulation peaks at approximately 4 radians at the time of merger.
The extrapolation error ($\phi_{ext}$) provides a constant error estimate at approximately 0.01 radians, two orders of magnitude smaller than the discretization error at the time of merger, and is the only significant error for the first few hundred $t/M$.
The error estimate from the loss of mass during the simulation is negligible, even at its maximum value, which occurs after $t_{\text{peak}}$ (the time of merger), indicated in the plots by the vertical dashed line.

\begin{figure}
  \includegraphics[width=\columnwidth]{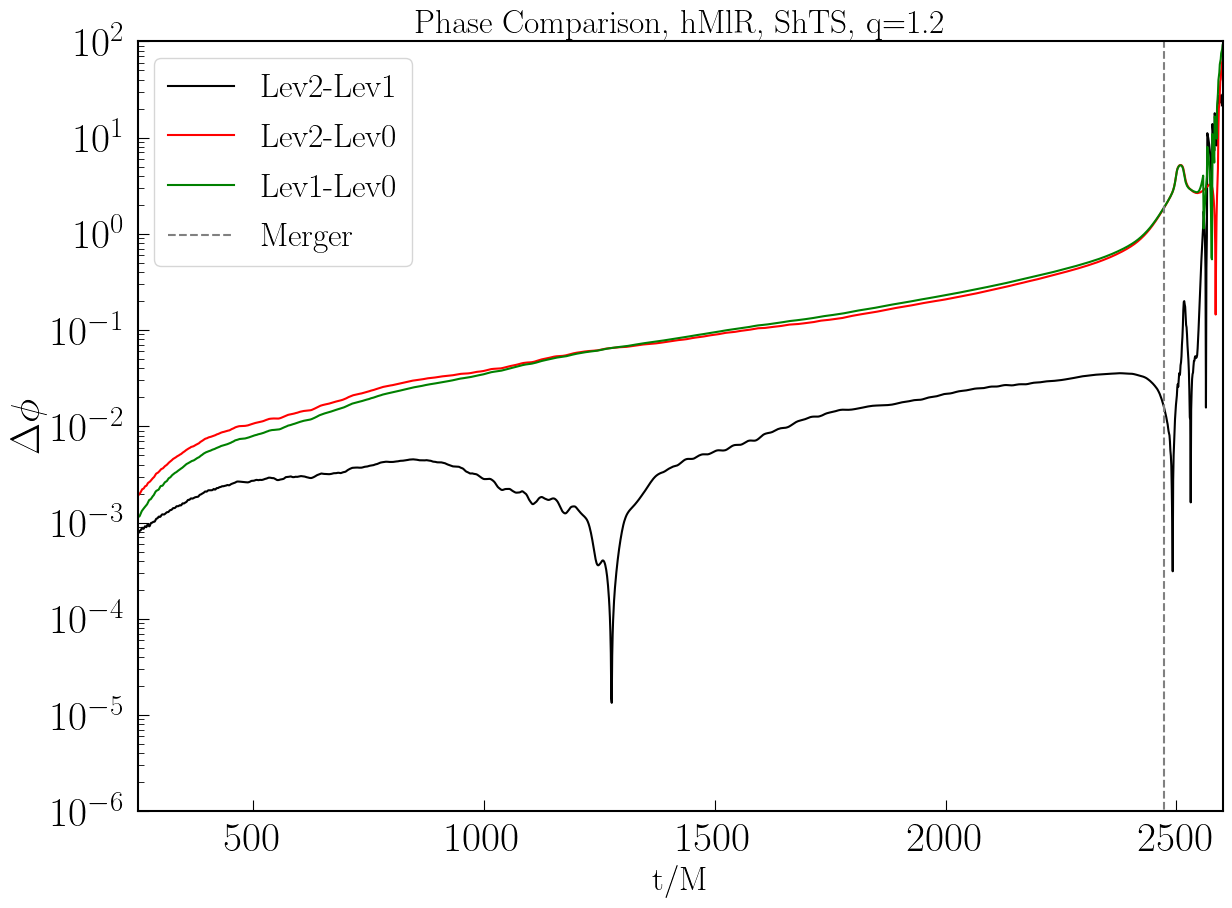}
  \caption{Comparison of phases for the extrapolated gravitational waves at three resolutions for the $q=1.2$ hMlR ShTS waveforms.
  There is minimal difference between the Lev1 and Lev2 phases during the simulation. 
  This is clearly due to cancellation of errors, with a $\Delta \phi=0$ crossing at $t\sim 1300M$ (a similar behavior is seen in the \qone1 case).}
  \label{fig:phasecomp_q12}
\end{figure}

\begin{figure}
  \includegraphics[width=\columnwidth]{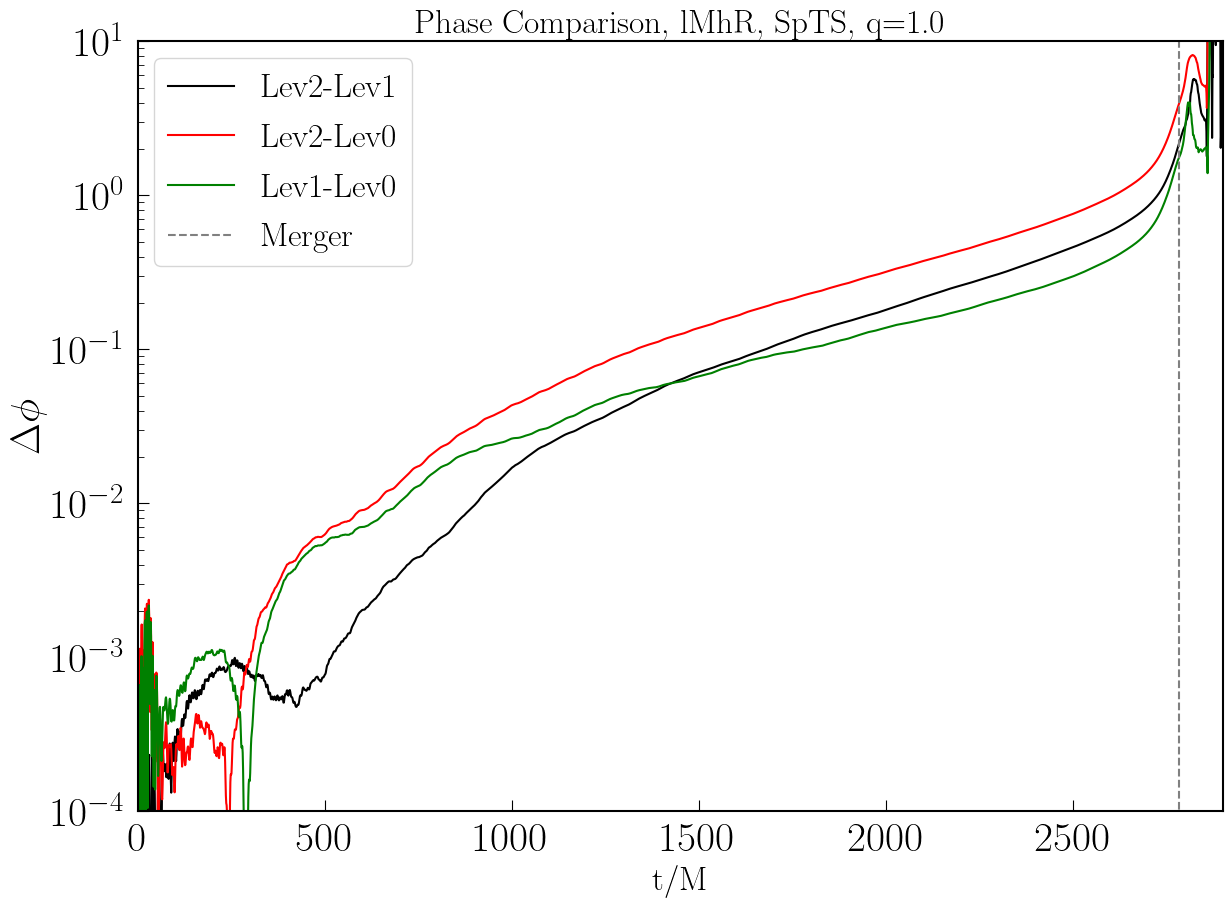}
  \caption{Difference of phases for the extrapolated gravitational waves of the lMhR SpTS $q=1.0$ waveforms.
   In this case, there appears to be partial cancellation of errors between the Lev0 and Lev1 simulations at late times.
  }
  \label{fig:phasecomp_q10_lMhR}
\end{figure}

In figure \ref{fig:phasecomp_q12}, we compare the phase difference between resolutions for the hMlR \qone2 systems.
We see a significant difference between the higher two resolutions (Lev1 and Lev2) and the low resolution (Lev0).
The Lev1 and Lev2 simulations are on the other hand very close to each other. 
This is most likely due to error cancellations between the two highest resolution as the error changes sign twice during the evolution.
We see similar behavior in the hMlR \qone1 waveform.
This is a fairly common issue when evaluating simulation errors with the SpEC code, and the main reason that any estimate of the phase error in the waveform requires three or more resolutions.

For the lMhR \qone0 case (figure \ref{fig:phasecomp_q10_lMhR}), we see a different behavior, as the Lev1 and Lev2 simulations minimally differ.
However, between approximately a quarter and halfway through the simulation, the Lev1 phase begins to drift towards the Lev0 simulation.
The accumulation of this error is likely due to the changing signs of some canceling errors, similar to the ShTS $q=1.2$ simulation.
As the simulation progresses, different errors dominate, and this figure is another example of the necessity of more than two resolutions.
This is a natural consequences of the resolution choices made in SpEC, where different sources of error (spectral, finite difference, time stepping) are kept at roughly the same level in order to avoid wasting computational resources on over-resolving one sector of our simulations -- but as a result may occasionally cancel out when comparing simulations at two resolutions only.

\begin{figure*}
  \includegraphics[width=0.45\linewidth]{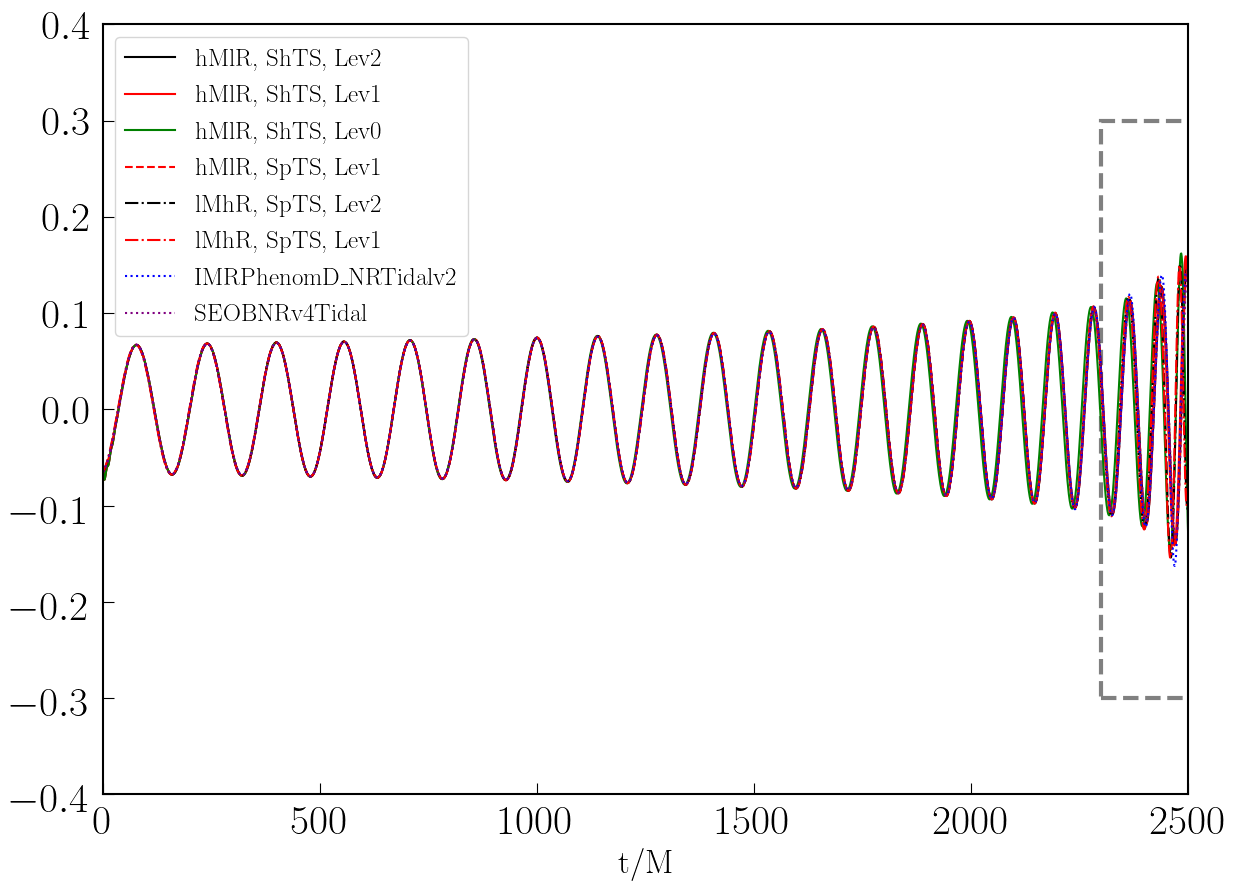}
  \includegraphics[width=0.45\linewidth]{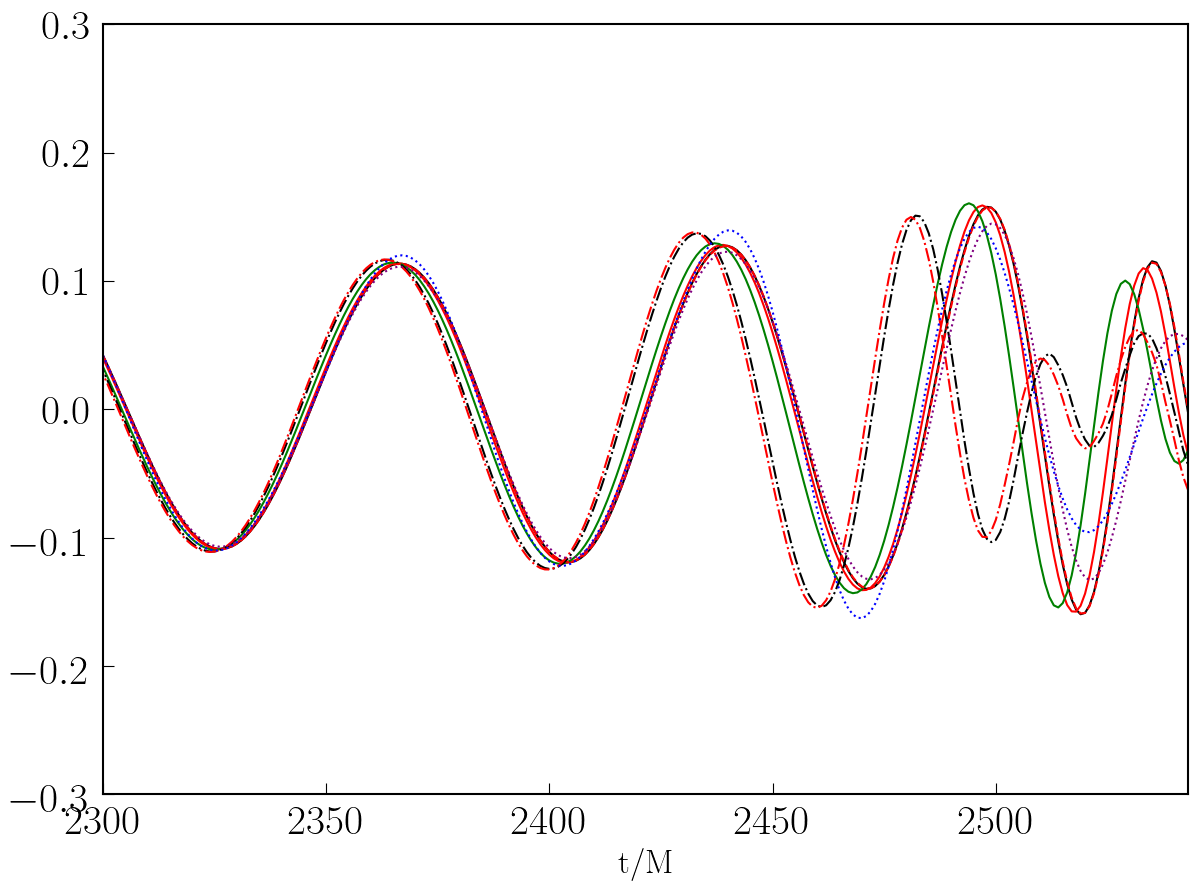}
  \caption{Gravitational wave signals for the \qone1 simulations. 
  We show three resolutions for the hMlR ShTS $q=1.1$ simulations, one from the hMlR SpTS \qone1, two from the lMhR SpTS \qone1 simulations, and two from the analytical models IMRPhenomD\_NRTidalv2 and SEOBNRv4 generated for the hMlR EoS.
  The dashed gray square indicates the zoomed in plot to the right.
  It can be seen at merger a clear distinction between the lMhR EoS and the hMlR EoS. Even the Lev0 system with a relatively large error is still visibly distinct from the lMhR systems.
  This demonstrates that we are able to distinguish between two closely-related EoS (numerically; this does not mean that such a difference is observable by current GW detectors).
  We can also see close matching to the analytical models until just before merger, when slighlty larger differences arise.}
  \label{fig:analytic-nr-comp-q11}

\end{figure*}
\begin{figure*}
  \includegraphics[width=0.45\linewidth]{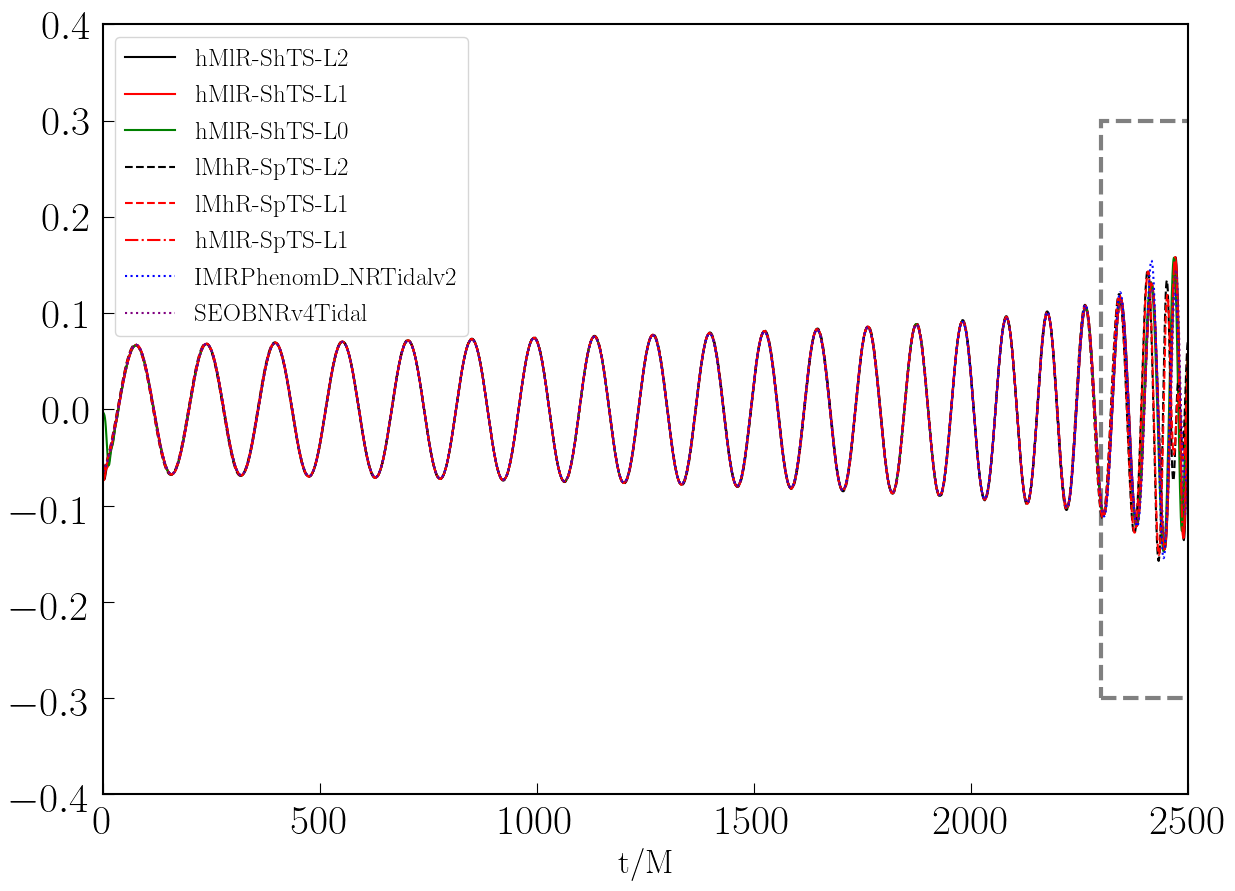}
  \includegraphics[width=0.45\linewidth]{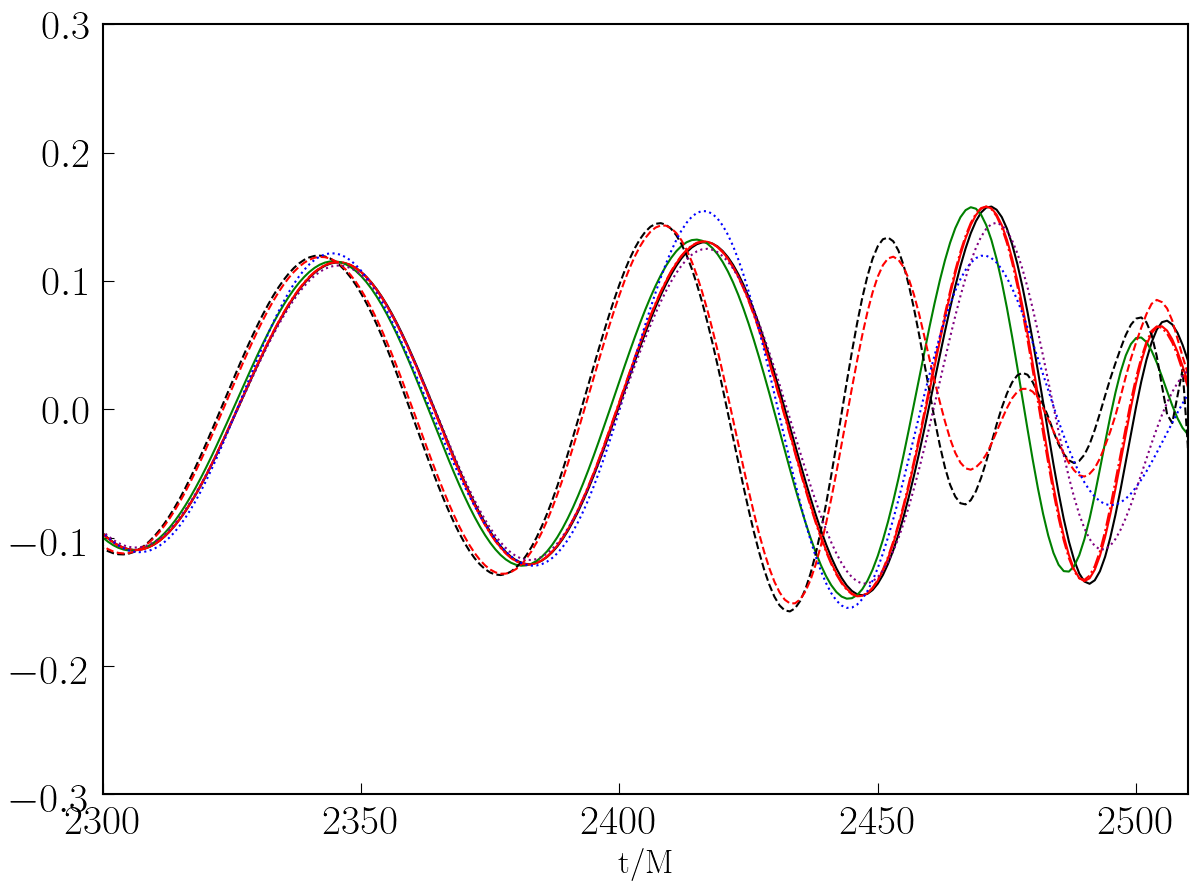}
  \caption{Gravitational wave signals for the \qone2 simulations. 
  We show three resolutions from the hMlR ShTS, one from the hMlR SpTS simulation, two from lMhR SpTS, and finally two from the analytical models IMRPhenomD\_NRTidalv2 and SEOBNRv4 generated for the hMlR EoS.
 We see a clear difference between the hMlR EoS and the lMhR EoS, as in the \qone1 case above. The analytical models also closely match our simulations, until merger.
  }
  \label{fig:analytic-nr-comp-q12}
\end{figure*}

\begin{figure*}
  \includegraphics[width=0.45\linewidth]{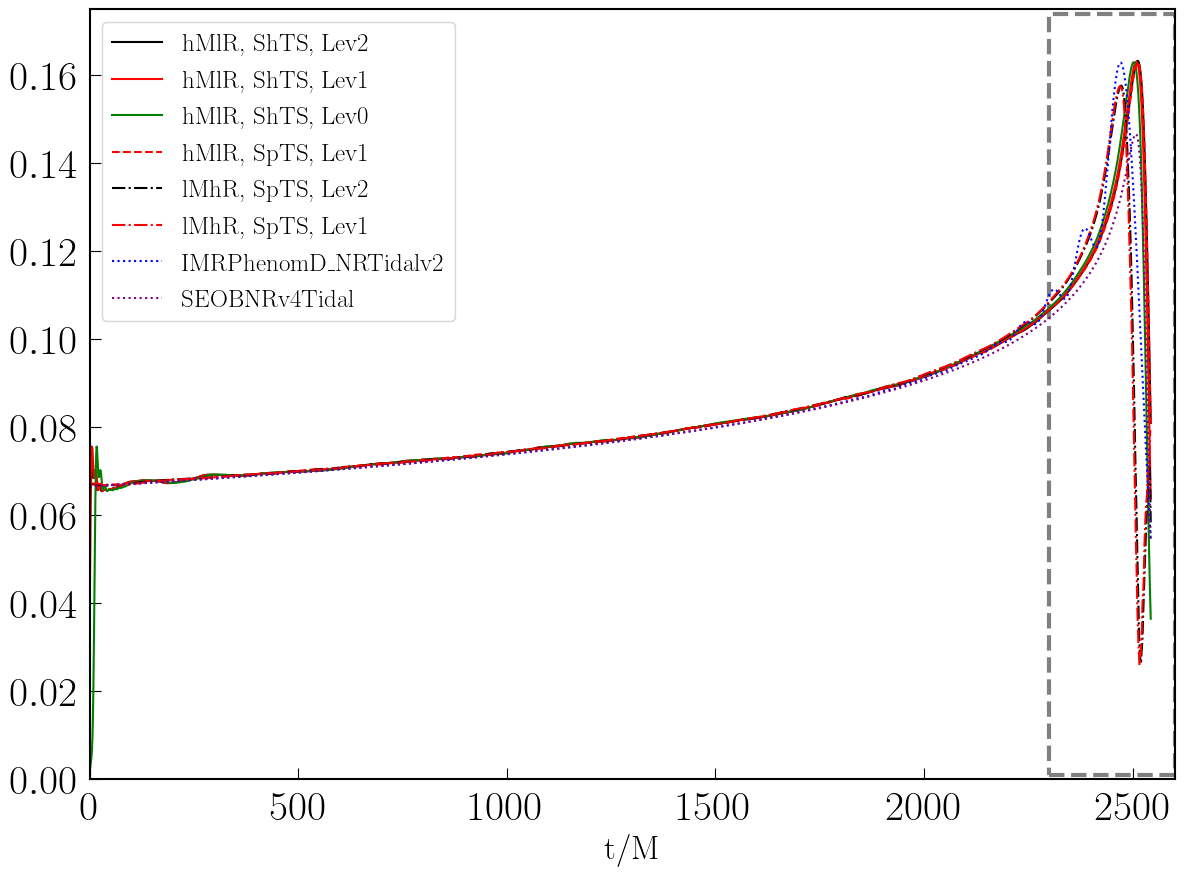}
  \includegraphics[width=0.45\linewidth]{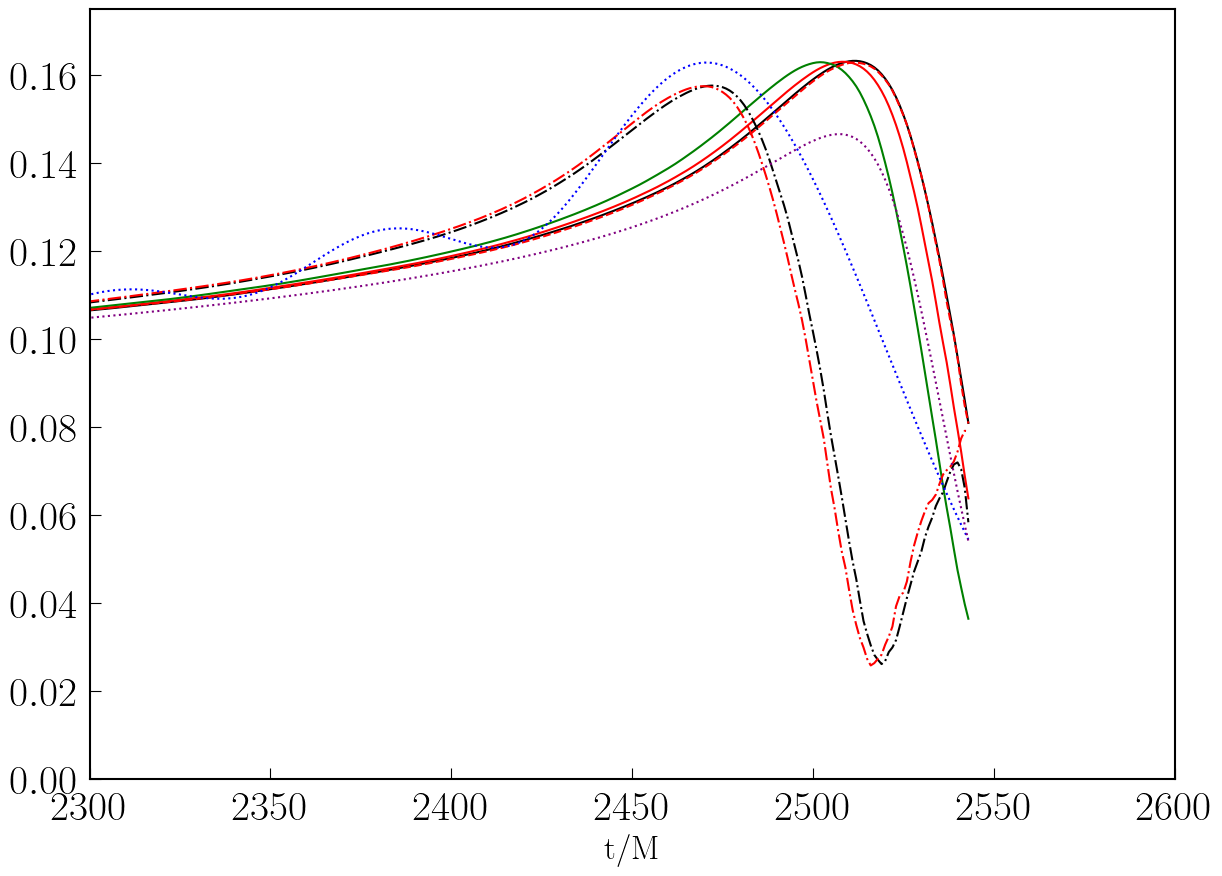}
  \caption{Amplitude of the gravitational waves of the hMlR ShTS \qone1 systems, the hMlR SpTS \qone1 system, two lMhR SpTS \qone1 simulations, and the \qone1 hMlR \seo and \imr models, after allowing for a time and phase shift minimizing differences at early times.
  The dashed gray square indicates the area covered by the plot to the right, which focuses on the merger.}
  \label{fig:analytic-nr-amplitude-q11}
\end{figure*}

\begin{figure*}
  \includegraphics[width=0.95\linewidth]{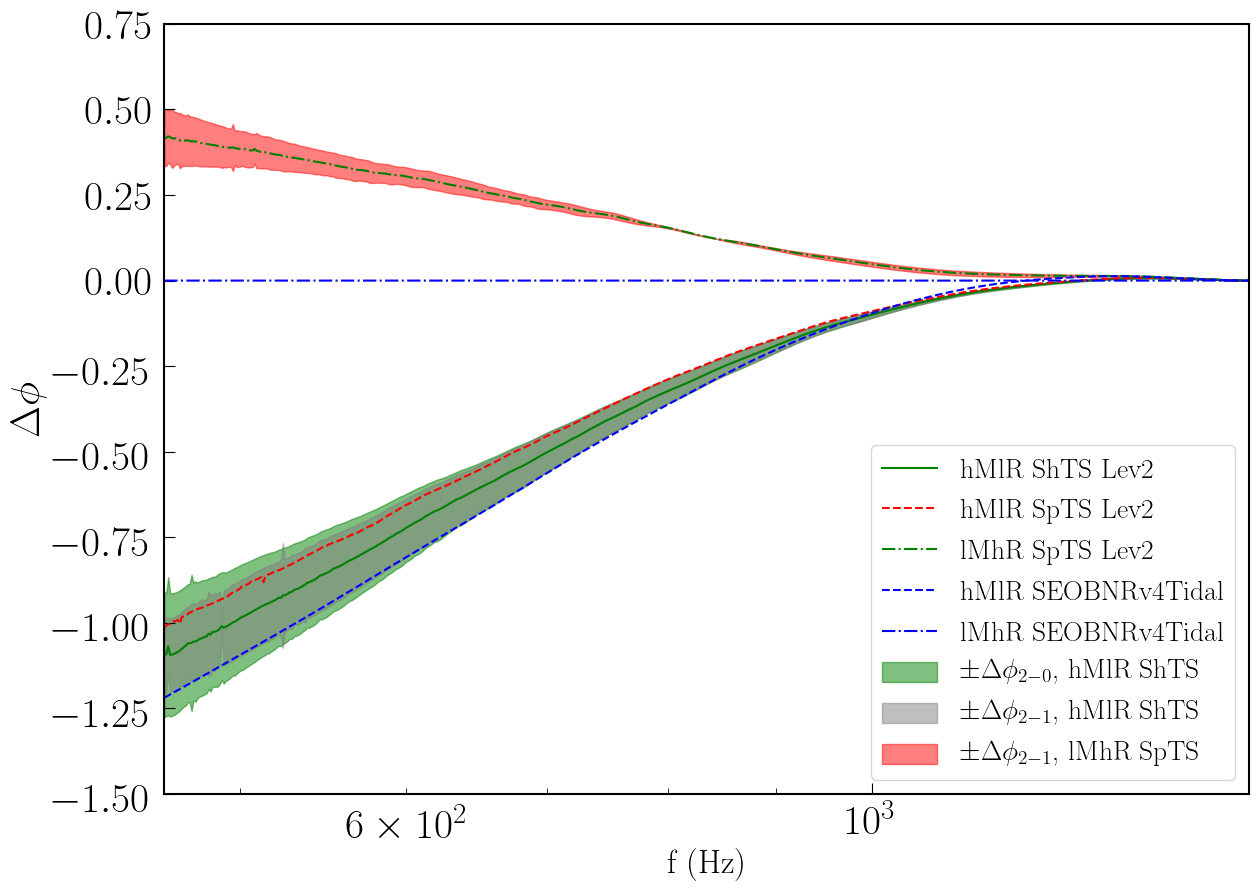}
  \caption{Phase error between the lMhR \seo analytical model and other simulations/models as a function of frequency.
  In this plot, we use the methods of analysis from~\cite{Read_2023}, equation 19b in particular.
  The filled-in areas are defined by $\Delta\phi_{2-1}=1.77|\phi_{\text{Lev2}}-\phi_{\text{Lev1}}|$ and $\Delta\phi_{2-0}=0.7|\phi_{\text{Lev2}}-\phi_{\text{Lev0}}|$, i.e. errors assuming second order convergence of the simulations.
  In this analysis, we choose as reference frequency $f_c=1514$ Hz, the lowest value of maximum frequency among our simulations and analytical models (the analysis effectively assumes that all simulations are at the same time and phase when reaching that frequency).
  }
  \label{fig:freq-comparison-q12}
\end{figure*}

\subsection{Discussion}

Given the relatively minimal database of similar BNS simulations, we compare the accuracy of our simulations to the BNS simulations in Foucart {\it et al} (2019)~\cite{Foucart_201902}.
Compared to~\cite{Foucart_201902}, we can see that our total error at merger is comparable to the \qone0 simulations performed with the piecewise polytropic EoS MS1b, and slightly higher than for the single polytrope with $\Gamma=2$ case.
However, the MS1b evolved for 8.5 orbits, shorter than our present simulations of about 10.5 for the ShTS and 11.5 for the SpTS, which themselves are shorter than the 12.5 orbits of the single polytrope simulation. 
From this, we can see that we accumulate error slower than the MS1b EoS, and faster than the polytropic EoS.

We additionally can directly compare our two ShTS systems against two analytical models, IMRPhenomD\_NRTidalv2 and SEOBNRv4Tidal, generated using the LALSuite~\cite{lalsuite}.
Both analytical models were generated using our simulations' parameters (ADM mass, tidal deformability, etc.) and show a high level of agreement in waveforms (see figures \ref{fig:analytic-nr-comp-q11} and \ref{fig:analytic-nr-comp-q12}, discussed below). 

In order to perform meaningful comparisons with analytical models and between simulations of different lengths or using different equations of state, the gravitational waves from our simulations are time and phase matched to the Lev2 resolution simulations by choosing two set times in the reference waveform, and minimizing the phase difference within that time frame among all transformations $t'=t+\delta t$, $\phi'=\phi+\delta\phi$.
Before doing so, in order to remove any difference in time steps between gravitational wave sets, we interpolate the highest resolution gravitational wave (referred from here as our 'target' wave) to a standardized time series using a cubic 1D interpolator.
For each gravitational wave we wish to match to our target wave (referred from here as the 'adjusted' wave) we go through the following procedure:
Each wave is interpolated to the standardized time series.
An allowed maximum time shift $\Delta t$ is chosen for the adjusted wave, and a minimizing algorithm then determines the proper time and phase shifts to reduce the phase error between the adjusted wave and the target wave within the chosen matching window.

We can see in figures \ref{fig:analytic-nr-comp-q11} and \ref{fig:analytic-nr-comp-q12} the analytical and numerical relativity waveforms overlaid, and, for waveforms using the same equation of state, the close agreement between them during the inspiral and plunge phases of the simulation. Differences between simulations and models are most apparent near merger, but this is not unexpected as analytical models often do not accurately predict the merger portion of simulations. Simulations using different equations of state are clearly distinguishable at merger on these plots.
Using the peak amplitude of the extrapolated and analytical waveforms as the merger time, we can see in figure \ref{fig:analytic-nr-amplitude-q11} that the \seo analytical model has an oscillatory amplitude before merger, and reaches merger before the other hMlR \qone1 systems.
We see similar behavior for the \qone2 systems.

As an additional avenue of waveform analysis, we use a new method from Read (2023)~\cite{Read_2023}.
Gravitational waves are Fourier transformed using the Stationary Phase Approximation, and matched in time and phase at a reference coalescence frequency $f_c$ (chosen to be the minimum peak frequency among our simulations and analytical models in our analysis). 
We then compare the resulting phase differences of the Fourier transform.
Quite importantly, the method is constructed so that the phase difference is fairly insensitive to numerical noise in the calculation of frequencies and of their derivatives, and relies on the choice of a single reference frequency for matching waveforms. 
In~\cite{Read_2023}, a variation of this method is also used to assess whether waveforms are distinguishable in various gravitational wave detectors -- however, this requires knowledge of the waveforms over the entire frequency range accessible to the detector, while our numerical waveforms only provide data for $f\gtrsim 400\,{\rm Hz}$. 
Another important advantage of this method is that it is relatively insensitive to errors in the early phase of the evolution, when simulations can have a hard time resolving high-frequency noise, and instead provide a more direct comparison of waveforms in the range in which finite size effects are the largest.

  We can see in figure \ref{fig:freq-comparison-q12} that the hMlR \seo model is within our numerical error. 
  This is not the case for the lMhR simulation. 
  This is due to the fact that, in this plot, we are matching waveforms at the peak frequency of the lMhR configuration, which is about $100\,{\rm Hz}$ below the peak frequency of the hMlR configuration. 
  A similar comparison matching waveform at the peak frequency of the hMlR configuration shows the \seo model now inconsistent with the numerical simulations. 
  This is another clear indication that any difference between the \seo model and the numerical result is due to the behavior of the \seo model close to merger.
  Additionally we see that our simulations can clearly distinguish between the hMlR and lMhR EoS.
  However, we must note that the difference between the two EoS mostly comes from the behavior of the waveform above 1 kHz, outside the sensitive range of LIGO and Virgo.
Accordingly, these results do not indicate anything about the detectability of these differences by current gravitational wave detectors. 
They only tell us that the numerical relativity waveforms are distinguishable well outside of their numerical errors. 
This can be better understood if we note that applying a time and phase shift on these waveforms allow us to change any curve on figure \ref{fig:freq-comparison-q12} according to $\phi \rightarrow \phi + Af + B$ for any constant $A,B$. 
Waveforms are thus only distinct on this plot if they differ in their second derivative $\frac{d^2\phi}{df^2}$. Clearly, this is only the case at high frequency ($f\gtrsim 900\,{\rm Hz}$).
  
Similarly, in figures \ref{fig:analytic-nr-comp-q11} and \ref{fig:analytic-nr-comp-q12}, we are clearly able to see a distinct phase difference between our hMlR EoS and the lMhR EoS within our current numerical error.
Assuming a linear dependence of the phase at merger in $\tilde \Lambda$, and for the specific EoS and mass ratios simulated in this manuscript, we estimate that we are able to distinguish with SpEC different gravitational waveforms down to a dimensionless tidal deformability difference of $\Delta \tilde \Lambda \approx 55$, well below current constraints from the observation of GW170817~\cite{Abbott_201710}. This shows that the spectral EoS is a promising option to train analytical models.

Moving to a comparison of the time stepping methods, we see a very close agreement in the extrapolated waveforms using the ShTS method and the SpTS method.
In both the \qone1 and \qone2 cases, the SpTS and ShTS simulations behave nearly identically in waveform and peak amplitude at merger, with differences well below our estimated numerical errors.

As the ShTS and SpTS simulations were run on different clusters, specifically the University of Texas' Frontera cluster and the University of New Hampshire's Plasma cluster, a direct comparison of computational cost is non-trivial.
We therefore compare two main components to estimate computational cost: the number of time steps the grid evolving Einstein's equations took and the CPU hours spent during the $\Delta t=5000$ (about 0.025 seconds) preceding merger.
Using this time period will avoid the initial numerical errors and junk radiation at simulation start from affecting the computational time, as well as the extra orbit in the SpTS case.
We compared the Lev1 \qone1 simulations.

We found the ShTS simulation to have taken 251823 steps during this $\Delta t=5000$ period, and cost 60341 CPU hours.
In comparison, the SpTS took 357506 steps, with only a cost of 42707 CPU hours. 
We also must note that the SpTS simulation had an approximately 10\% increase in resolution compared to the ShTS case, which should result, everything else being equal, in approximately 40\% increased computational expense for the fluid evolution.
Despite this additional cost and approximately 42\% additional time steps for the evolution of Einstein's equations, we can see an approximately 30\% reduction in computational time cost in the SpTS case.
From standardized speed tests performed on both machines for BNS evolutions, we can also determine that Frontera is roughly 12\% faster than Plasma, further increasing the CPU hour cost reduction of the SpTS method.
While this comparison is relatively rough, we find it sufficient to state that the SpTS method does save on computational resources in simulations such as the ones used in this paper.

\section{Conclusion}

From our work, we have found the spectral EoS to be a promising option for numerical BHNS or BNS waveform studies.
It is capable in our systems to generate gravitational waves from BNS that agree within expected error with state-of-the-art analytical models up to the merger event, where analytical models become less accurate.
The defining parameters of spectral EoS can be adjusted to produce a range of neutron star EoS candidates, offering a large amount of flexibility for future systems as we further refine the constraints on the EoS of a neutron star.
It offers an improved ability to generate stars with appropriate macroscopic properties when compared with a polytropic EoS, and provides better numerical accuracy compared to a discontinuous EoS, at least in the SpEC code.

We find in particular that, for two distinct methods of matching the waveforms in time and phase, we are capable of clearly capturing differences in the gravitational wave signals produced by binaries with tidal deformabilities of $\tilde \Lambda \approx 550$ and $\tilde \Lambda = 700$. 
Assuming a linear dependence of the phase differences with $\tilde \Lambda$, our results indicate that variations of $\Delta \tilde \Lambda \approx 50$ could lead to numerical waveforms whose behavior close to merger differ by more than our current finite-resolution errors.

As for numerical methods, our preliminary comparison between the SpTS and ShTS methods indicates potential computational cost savings by uncoupling the finite difference and spectral grid time steps.
In this manuscript, we measured a greater than 30\% decrease in CPU hours used for a simulation with $\approx 10\%$ increased resolution on a cluster with slower hardware.
Clearly, further testing on simulations conducted on the same cluster, utilizing the same simulation parameters such as resolution and initial conditions is required for a definitive answer, but our test here has shown the comparison to be worth closer inspection of a method that could potentially have significant savings in computational cost.

There is still a great deal of experimentation that can be done with the spectral EoS, including higher resolution simulations to verify the accuracy of our current error estimates.
Additionally, generating NS with varying radii, mass, and tidal deformability by adjusting the $\Gamma_0$, $\eta_2$, $\eta_3$, $\Gamma_{\text{th}}$, $\rho_0$ and $P_0$ may prove useful in determining its viability in a range of systems, allowing for a smooth EoS for SpEC and other codes sensitive to discontinuous EoS.
The spectral EoS offers a new avenue for simulations, expanding our potential tools for more accurate and better resolved simulations, to aid in eventually better understanding the detected gravitational waves from merger events between compact objects.

\acknowledgements

A.K and F.F. gratefully acknowledge support from the Department of Energy, Office of Science, Office of Nuclear Physics, under contract number
DE-AC02-05CH11231 and from the NSF through grant AST-2107932. M.D. gratefully acknowledges support from the NSF through grant PHY-2110287.  M.D. and F.F. gratefully acknowledge support from NASA through grant 80NSSC22K0719. M.S. acknowledges funding from the Sherman Fairchild Foundation
and by NSF Grants No. PHY-1708212, No. PHY-1708213, and No. OAC-1931266
at Caltech.  L.K. acknowledges funding from the Sherman Fairchild Foundation
and by NSF Grants No. PHY-1912081, No. PHY-2207342, and No. OAC-1931280
at Cornell. Computations for this manuscript were performed on the Plasma cluster, a Cray CS500 supercomputer at UNH supported by the NSF MRI program under grant AGS-1919310, and on the Wheeler cluster at Caltech, supported by the Sherman Fairchild Foundation. The authors acknowledge the Texas Advanced Computing Center (TACC) at The University of Texas at Austin and the NSF for providing resources on the Frontera cluster~\cite{10.1145/3311790.3396656} that have contributed to the research results reported within this paper. Computations were also performed on ACCESS resources through grant No PHY990002.

\bibliography{mybib}
\bibliographystyle{ieeetr}
\end{document}